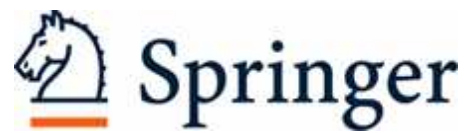

# Archaea-based Microbial Fuel Cell Operating at High Ionic Strength Conditions

# Archaea-based Microbial Fuel Cell Operating at High Ionic Strength Conditions

Ximena C. Abrevaya[a], Natalia Sacco[b], Pablo J.D. Mauas[a] and Eduardo Cortón[b]*

[a]University of Buenos Aires-CONICET, Instituto de Astronomía y Física del Espacio (IAFE). Ciudad Universitaria, (C1428EGA), Buenos Aires, Argentina.

[b]University of Buenos Aires. Facultad de Ciencias Exactas y Naturales y CONICET, Department of Biochemistry, Buenos Aires, Argentina.

*Corresponding author: Eduardo Cortón, Fax: INTL + (54-11) 4576-3342, Phone: INTL + (54-11) 4576-3343/78/79 Ext. 203, e-mail: eduardo@qb.fcen.uba.ar

**ABSTRACT**

In this work two archaea microorganisms (*Haloferax volcanii* and *Natrialba magadii*) used as biocatalyst at a microbial fuel cell (MFC) anode were evaluated. Both archaea are able to grow at high salt concentrations. By increasing the media conductivity, the internal resistance was diminished, improving the MFCs performance. Without any added redox mediator, maximum power ($P_{max}$) and current at $P_{max}$ were 11.87 / 4.57 / 0.12 µW cm$^{-2}$ and 49.67 / 22.03 / 0.59 µA cm$^{-2}$ for *H. volcanii*, *N. magadii* and *E. coli*, respectively. When neutral red was used as redox mediator, $P_{max}$ was 50.98 and 5.39 µW cm$^{-2}$ for *H. volcanii* and *N. magadii* respectively. In this paper an archaea MFC is described and compared with other MFC systems; the high salt concentration assayed here, comparable with that used in Pt-catalyzed alkaline hydrogen fuel cells will open new options when MFC scaling-up is the objective, necessary for practical applications.

# 1. Introduction

A microbial fuel cell (MFC) is a device that converts chemical energy stored in organic substances or other reduced compounds into electrical energy by using microorganisms as biocatalysts. MFCs have had a long history in the academic world, from the first description of the phenomena by Michael Cresse Potter (1911), when he placed a platinum electrode into cultures of yeast or *E. coli* and showed that a potential difference was generated. Later, MFCs were rediscovered by Benetto (Benetto, 1984; Allen and Benneto, 1993).

Currently, there are several factors limiting the performance of MFCs and inhibiting the progress of applying MFCs in practice. These limiting factors include the activity of biocatalysts (microorganisms), electrodic reactions (both in cathode and anode), internal resistance and reactor design, among others. In the last 10 years, a new paradigm about extracellular electron transfer without the assistance of extracellular (added or produced by bacteria) redox mediators (as flavins) have raise into consideration new limiting factors and mechanisms. Direct electron transfer via outer-surface c-type cytochromes, long-range electron transfer via microbial nanowires and electron flow through a conductive biofilm matrix containning cytochrome have been proposed (Baron et al, 2009; Lovley DR, 2011). Besides, the chemical and physical working conditions of MFCs (at least at the bio-anode compartment) are dictated by the nature of the biological component, a key element of MFCs. Depending on the microorganism and growth conditions, changes in the external chemical and physical conditions can bring about alterations in several primary physiological parameters, inhibiting growth and metabolism, and eventually causing the death of the microorganisms used as biocatalysts at the anode.

In the last years, several MFCs designs have been assayed, in which multiple combinations of electrode material, microorganism and other parameters were evaluated, as discussed in recently published reviews (Rabaey and Verstraete, 2005; Bullen et al., 2006; Davis and Higson, 2007; Debabov, 2008; Osman et al., 2010; Lovley, 2011).

A previous study (Jang et al., 2004) showed that an MFC could be operated increasing the salt concentration at the cathode compartment, without affecting the survival of the microbial communities at anode. The study showed that when NaCl concentration at the cathode was increased from 0.1 to 1 M, the produced current increased from 3.5 up to 7.7 mA. Also, Liu et al. (2005) proved that by increasing the anode ionic strength (IS) from 100 to 400 mM of NaCl, the internal resistance ($R_{int}$) was lowered and the maximum power density was increased. However, electricity production at MFCs, as far as we know, has been only previously linked to the metabolic activity of only very few extremophiles, salt-tolerant microorganisms. Arsenate respiring bacteria isolated from moderately hypersaline Mono Lake (*Bacillus selenitireducens*) was previously used, with a maximum power of 18.5 mW m$^{-2}$ (Miller and Oremland, 2008). ~~Also, an archaea-based MFC was suggested, as a probe of concept for a device, based on detection of metabolic reductive process, presented as a life-searching device (Abrevaya et al., 2010).~~

The objective of this research was to investigate the performance (by means of polarization curves) of halophile archaea used at MFC anode and the effect of high salt concentration both at anode and cathode compartments. We hypothesize that such increase at the media conductivity would improve the MFCs performance (electricity production) by reducing $R_{int}$, among other possible factors. To achieve this, two halophilic archaea were used and the results obtained were compared with a bacterial strain, using an identical hardware (*H. volcanii, N. magadii* and *E. coli*, respectively). *E.*

*coli* was used previously for other authors in mediated and non-mediated MFCs (Ieropoulos et al., 2005) and it is worthy of comparison studies.

We evaluate these microorganisms at non-mediated (more precisely, non-added mediator) and at mediated MFCs, where a mediator (neutral red, NR) was used to improve electron transport between the oxidative microbial metabolism and the anode surface. Also, the use of NR allows us to compare the different MFCs in a condition where the current is not limited by naturally occurring mediators or mediator-like substances that could be presented in the culture media produced by the microbial strains used. NR was used because it has a redox potential of -325 mV, similar to that of NADH (-320 mV vs. SHE), and a structure similar to that of flavins. Its redox potential suggests that NR could interact with metabolic steps prior to respiratory chains (McKinlay and Zeikus, 2004). MFCs described here are based on plain Toray® carbon paper electrodes, and Nafion® was used as a proton transporter membrane.

We show in this work that an archaea microorganism can be used as biocatalyst in MFCs, and electricity generation is possible. Furthermore, the effect of high conductivity in both current production and internal resistance is shown. Previously published work where biofilm-forming electrogenic bacteria was used, show serious scaling-up limitations. Here we open new possibilities for the design and operation of MFCs.

The new and amazing possibilities of *H. volcanii* and other extreme microbial physiologies could be a key to increase the maximum current density and power obtained with MFCs. The use of an added mediator allowed us to compare both ionic strength conditions; for potential applications many naturally occurring or microbially synthesized compounds can serve as electron carriers.

## 2. Experimental

### 2.1. Microbial strains and microbiological methods

*H. volcanii* strain DS70 (DS2 cured of pHV2, Wendoloski et al., 2001) was grown aerobically at 35 °C, until an OD (600 nm) of ca. 1 was reached. Growth medium Hv-YPC contains (g L$^{-1}$), yeast extract (5), peptone (1), casaminoacids (1), NaCl (144), $MgSO_4 \cdot 7H_2O$ (21), $MgCl \cdot 6H_2O$ (18), KCl (4.2), $CaCl_2$ (0.35), and Tris-HCl (1.9); pH was adjusted to 7.0.

*N. magadii* (ATCC 43099) was grown aerobically at 37 °C with shaking at 200 rpm. Growth medium composition was (g L$^{-1}$): yeast extract (5), NaCl (200), $Na_2CO_3$ (18.5), Sodium citrate (3), KCl (2), $MgSO_4 \cdot 7H_2O$ (1), $MnCl_2 \cdot 4H_2O$ (3.6 x 10$^{-4}$), $FeSO_4 \cdot 7H_2O$ (5 x 10$^{-3}$), with pH adjusted to 10 (modified from Tindall et al., 1984). The optical density of the cultures was spectrophotometrically measured at λ=600 nm until it reached an OD of ca. 0.5.

*Escherichia coli* K-12 derived strain (ATCC 15153) was used. It was grown aerobically at 37 ºC until an OD (660 nm) of ca. 1.0 is reached. Tryptic Soy Broth (DIFCO) was used as culture media, dissolving 30 g in 1 L of distilled water, and adjusted to pH 7.2 if necessary. The broth prepared in this way contains (g L$^{-1}$), pancreatic digest of casein (17), enzymatic digest of soybean meal (3), NaCl (5), $K_2HPO_4$ (2.5) and glucose (2.5).

## 2.2. Microbial fuel cell hardware

The MFC used had two compartments with a volume of one-liter, separated by 4 cm$^2$ of Nafion® 115 membrane (from FuelCellStore, San Diego, CA). The cell was made on 6 mm thick transparent acrylic, and had a lid with six 0.6 cm diameter holes, allowing connections for electrodes, gas bubbling and sample removal / reagent addition. Before inoculation with the microbial biocatalyst, the anolyte solution was purged with $N_2$ for 20 min in order to remove oxygen. The catholyte solution was bubbled continuously with air to allow mixing. Electrode separation was 2 cm. Before its use, the cell was sterilized by immersion overnight into 10 % v/v $H_2O_2$, followed by distilled (double osmosis) sterile water rinse.

## 2.3. Electrodes

MFC cathode and anode were made of plain carbon paper TGP-H-030 (Toray®, Tacoma, WA), with a density of 0.40 g cm$^{-3}$ and a porosity of 80%. The geometrical area of anodes and cathodes was 10 cm$^2$. Before their use, they were cleaned by consecutive immersion during 1 h in 1 mol L$^{-1}$ HCl and NaOH, rinsed exhaustively and stored in distilled sterile water.

## 2.4. Catholyte and anolyte composition

*Haloferax volcanii* anolyte was Hv-YPC growth medium. In some experiments, final concentration of 0.1 mM in NR was used as redox mediator. Ionic strength (IS) was ca. 2.68 M. Catholyte was modified *H. volcanii* growth medium Hv-YPC; yeast extract, peptone and casaminoacids were not included; it contained ferricyanide (8.4 g L$^{-1}$). IS was ca. 2.72 M.

*Natrialba magadii* anolyte was *N. magadii* grown medium. In some experiments, final concentration of 0.1 mM in NR was used as mediator. IS was ca. 3.63 M. Catholyte was modified *N. magadii* growth medium; yeast extract was not included. Ferricyanide concentration was 8.4 g $L^{-1}$. IS was ca. 3.67 M.

*E. coli* anolyte contained (g $L^{-1}$), glucose (5), $Na_2HPO_4$ (6), yeast extract (5), $KH_2PO_4$ (3), $NH_4Cl$ (1), NaCl (0.5), $MgSO_4 \cdot 7H_2O$ (0.12), and $CaCl_2$ (0.01) dissolved in distilled water, pH adjusted to 7.0. In some experiments, final concentration of 0.1 mM in neutral red (NR) was used as mediator. Ionic strength was ca. 92.0 mM. Catholyte contains (g $L^{-1}$), $Na_2HPO_4$ (6), $KH_2PO_4$ (3), NaCl (0.5) and ferricyanide (8.4); pH was adjusted to 7.0. IS was ca. 125.0 mM.

**2.5. MFC set-up and operation**

The MFC was operated at 37 ºC with continuous air purging (cathode) to provide agitation. $N_2$ purging (anode) was used during measurements and 20 min before measurements or start-up. After purging, $N_2$ was used to provide agitation; no mechanical or magnetic stirring was used. The pellet obtained by centrifugation of 400 (*E. coli* and *H. volcanii*) or 800 (*N. magadii*) mL culture was used as inoculum (start-up) at anode compartment. The inoculated MFCs were allowed to stabilize overnight, with an external resistor (RL) of 4.7 kΩ, used in continuous operation. Before doing a voltammetric or polarization curve studies, the RL was disconnected 6 h to allow the system to reach open circuit (OC, without any external resistor) potentials. The experiments were usually carried out for 7 days. The polarization curves presented here represent typical experiments acquired when the OC and $P_{max}$ reach stable and elevated (plateau) values.

## 2.6. MFC analysis

In order to study the behavior of the MFC and its electrical characteristics, measurements were carried out, intercalating different external resistors (RL). Polarization curves were performed using RLs from 100 kΩ to 2.3 Ω; the potential E was measured by a digital tester with PC recording capabilities (Fluke 289). Current (I) production was calculated using the Ohm's law (I=V R$^{-1}$), where V is the voltage and R the resistance. Current density, $j$ (A cm$^{-2}$), was calculated as $j$=I S$^{-1}$, where S is the geometrical (projected) surface area of the anode electrode. Power density, P (W cm$^{-2}$), was calculated as P= I V S$^{-1}$. Internal resistance (R$_{int}$) was calculated from the slope of plots of V and I, using V=E$_{cell}$ - I R$_{int}$, where E$_{cell}$ is the electromotive force of the cell (Logan, 2008). We eliminate the data from the regions I and III (where polarization behavior is dominated by activation potential and mass transfer overpotentials) to construct quasi-linear plots to compute R$_{int}$.

## 2.7. Electrochemical studies

Cyclic voltammetry at low scan rate (1 mV s$^{-1}$) using a standard three electrode system was used to search for possible redox mediators at *H. volcanii* culture media or microbial culture; phosphate buffer (100 mM, pH 7) was used so as to investigate possible redox substances at Toray paper electrode. The window potential applied was from -400 to +500 mV; in order to obtain voltammograms in static conditions, N$_2$ was bubbled for only 10 min; following another 10 min (quiet time) the CVs were initiated.

A plain Toray paper anode was used as working electrode (WE), stainless steel wire as counter electrode (CE) and Ag/AgCl (KCl saturated) as reference electrode (RE). In order to investigate the NR reaction at the Toray electrodes, at the high IS used and in presence of *H. volcanii*, CVs between -700 and +700 mV were done. $N_2$ was used to remove oxygen; CVs at 50, 100, 200 and 400 mV s$^{-1}$ were made. All CVs were carried out in quiet solutions; before the beginning of the experiment, the WEs were poised for 1 min at the initial potentials. We used a potentiostat TEQ 03 (Ing. Sobral, La Plata, Argentina), with data acquisition and control *via* proprietary software.

## 3. Results and Discussion

### 3.1. Polarization and power curves

Figure 1 shows typical polarization and power curves for two of the three strains evaluated; we present four independent experiments with each strain (different inoculations) for *H. volcanii* and *E. coli* with and without an added redox mediator (neutral red). Numerical results can be observed at Table 1. When *H. volcanii* was used as anodic biocatalizer maximum power density ($P_{max}$) without added mediator was $11.87 \pm 0.54$ µW cm$^{-2}$, almost a 100 fold increase with respect to *E. coli*. We applied an independent two-sample t-Test to compare the $P_{max}$ obtained with both microbial strains, which provides evidence that the means are significantly different at $p \leq 0.01$, showing that the two strains produce different power output. The same trend was observed when current density (*j*) was compared, and values of $0.58 \pm 0.16$ and $49.67 \pm 0.81$ mA cm$^{-2}$ were obtained for *E. coli* and *H. volcanii*, respectively; *j* values presented are the ones obtained at $P_{max}$, as customary; data presented are averaged $\pm SD_{(n-1)}$.

Also in Table 1, we show the effect that NR addition has. Here the maximum power increases approximately five times in both *E. coli* and *H. volcanii* MFCs, showing that charge transport between microbial cells and electrodes could be improved by an external mediator

Furthermore, in Table 1 our results are compared to the ones published by other authors; our results show that P and I produced by our mediated archaea-based system was exceptionally superior to other mediated systems, reaching values approaching the highest standards established recently by non-mediated biofilm-based MFCs (Ishii et al., 2008). Although our non-mediated MFCs could be comparable concerning their electrical performance with other MFCs designs (Table 1), several critical factors (geometrical design, electrode size, membrane, etc.) forbid a direct comparison with other published data. However, when comparisons are made using the results obtained in this work, by means of the same MFC hardware, valid conclusion can be elaborated. It must be noted that $R_{int}$ (one important performance limiting factor) is strongly influenced by MFC set-up, geometry, electrodes, etc, and that the data presented by other authors are only partially comparable.

Some authors have obtained very interesting results for axenic cultures; Rabaey et al., 2005, reported a maximum of 88 mW $m^{-2}$ for *P. aeruginosa* without added mediator (this bacterium produces pyocyanin, which in this work is reported to be necessary for efficient electron transfer). Also, a maximum of 91 mW $m^{-2}$ for *E. coli* with redox mediator has been reported (Park et al. 2003). However, we must note that the comparison between different MFC set-ups is only partially accurate. *E. coli* in the set-up used by Park et al. produces about 19 times more power that in ours, so we can assume that this difference is related mainly to the set-up, the effect of a $Mn^{4+}$-graphite anode and that a $Fe^{3+}$-graphite cathode could be in part responsible for that. We hypothesize that, in such optimized MFC, *H. volcanii* could produce higher power densities.

We show that by adding a soluble mediator to halophile archaea (H. volcanii) MFCs performance is improved almost a 100 times with respect to a non-halophilic bacterium (*E. coli*). The co-culture of mediator producing halotolerant/halophile bacteria and halophile archaea (or other combinations) could produce a MFC system competitive with regard to the biofilm based ones.

Complementary work was done using an haloalkaliphilic archaea at extreme pH (pH = 10), *N. magadii* (Figure 2 and Table 1). In this experiment, 800 mL of *N. magadii* culture was used as anodic microbial suspension; polarization curves were measured after 2, 24 and 48 h. Then an external mediator was added at 0.1 mM final concentration, and a final polarization curve was measured after 4 h. The absorbance of the anodic microbial suspension was 0.407, 0.479 and 0.547 for 2, 24 and 48 h, respectively; the experiment show that $P_{max}$ increased from 1.01 to 1.92 and 4.57 µW cm$^{-2}$ as incubation time and absorbance increased. The increase in current and power production is likely related to the metabolic behavior of the archaea cells, given that the absorbance increased only slightly. The addition of an external mediator in this alkaline system (pH =10) slightly increased the values obtained previously without redox mediator. Using this system, it could be possible that other current limitations occur (as proton availability) which are stronger limitation to current production; also the differential membrane/metabolic characteristics of *N. magadii* could be responsible for this effect. Remarkably, the power and current production in *N. magadii*, although it was the highest IS used in this work, was (without NR) ca. half of the obtained with *H. volcanii*. This effect could be related to the different apparent ionic mobility of H$^+$ in water, which is about 6–7 times more than that of Na$^+$, (Wraight, 2006). Given the alkaline pH used in this experiment (pH = 10) and the high NaCl concentration (200 g L$^{-1}$), Na$^+$ must be important in charge transfer processes both inside and outside the Nafion membrane.

Here we compared three microbial strains using an identical hardware, demonstrating that the power and current were increased, and $R_{int}$ decreased when high IS and a halophile microorganism were used. The use of halophilic archaea as MFC anode biocatalyst improves the current and power. This is probably an effect of the lower internal resistance of the halophilic, no mediator-added MFC based on *H. volcanii* (ca. 447 $\Omega$).

Probably these effects mainly depend on the anolyte and catholyte IS, and are poorly related to the microbial physiology of the microbial strains used, given the low growth and metabolic rates of *H. volcanii*, when compared with *E. coli*. Results obtained with *N. magadii* also show a positive effect of increased IS, but a negative effect of alkaline pH (pH = 10) are possibly related to the different mobility of the ionic charge transporters at the anodic microbial suspension. However, more detailed experiments involving other halophile archaea are necessary to probe this hypothesis. When mediated systems were compared, the effect of NR was important with *H. volcanii* and less notorious for *N. magadii* cells; it is possible that NR works better as electron shutter for *H. volcanii* than for *N. magadii*, but we do not have enough data to speculate about this phenomenon. Therefore, experiments involving other documented microbial mediators (methylene blue, humic acid) could be also worthy to understand the difference between the two halophilic archaea assayed here.

### 3.2. Internal resistance

Internal resistance ($R_{int}$) is a key performance driver of fuel cells (Barbir, 2005). In mediated MFCs, ohmic resistance ($R_\Omega$) is usually the most important contributing factor to $R_{int}$. The three sources of ohmic voltage loss are: (a) resistance to ion migration within the electrolyte, (b) resistance to electron transport within the fuel cell components (electrodes, gas diffusion layer, current collectors), and (c) contact resistances (Logan, 2008). The salt concentration we used was comparable with alkaline fuel cells, were 6.6 M KOH is habitually used (Burchardt et al., 2002), allowing very low $R_{int}$ (less than 1 Ω cm$^{-2}$). The increase of NaCl is generally used at the electrolyte to improve the mass transfer of charged particles (Gil et al., 2003). The increase in the fuel cell performance a *H. volcanii* seems to be related to the increased mass transfer of charge transporters and to the increased proton availability in the cathode (pH decrease from 7.0 to 5.9 in *H. volcanni* MFC); using marine water and sediment, MFCs were reported to have better performance when compared to wastewater based ones (low IS), as reported previously (Tender et al., 2002; Bond et al., 2002). The effect of increased ionic strength was also assayed by Liu (Liu et al., 2005), where the ionic strength was increased from 100 to 400 mM, showing a noticeable power increase at high IS.

Electricity production at MFCs has been only previously linked to the metabolic activity of only very few salt-tolerant bacteria (Miller and Oremland, 2008), by using arsenate respiring bacteria isolated from moderately hypersaline Mono Lake (ML, *Bacillus selenitireducens*), and salt-saturated Searles Lake, CA (SL, strain SLAS-1); when pure culture bacteria were used, very low current was obtained for both strains, 49 and 59 µW m$^{-2}$, respectively. When the bacteria were assayed at MFC together with lake sediment, which could have some natural occurring redox mediator, significantly more power was produced. Also, in other experiments, they show that MFCs with ML sediment more power is produced (18.5 mW m$^{-2}$, salinity 90 g L$^{-1}$) than with SL sediment (1.2 µW m$^{-2}$, salinity 346 g L$^{-1}$). Although these results appear not to be consistent with our hypothesis (high power production at high salinity/IS), the highest power production at Mono lake is consistent with the following facts: microbial activities were greater in Mono Lake, *B. selenitireducens* grows faster than strain SLAS-1, and the presence of inorganic electron donors, especially sulfide, in Mono Lake sediment. Also, the presence in ML of a wider range of anaerobic bacteria capable of efficiently transferring electrons to the anode could be possible.

The physical design of non-mediated, biofilm-based MFCs, where the distance between bioelectrochemical reactions and the anode is minimal (these reactions occurring mainly at the biofilm layer) allow lower $R_{int}$ and higher currents. At the high salt concentrations used here we achieved non-mediated (non-added mediator) and mediated MFCs with $R_{int}$ comparable with the previously published work (Table 1). Our MFCs are expected to have relatively high $R_{int}$, considering the distance between electrodes and the presence of a Nafion membrane; high IS and neutral or acidic pH show an effective way to improve MFC performance, lowering $R_{int}$ considerably; also, the incorporation of NR have the same effect, at normal or high IS (Table 1). At pH 10, the effect over $R_{int}$ was less notorious, perhaps related to the lower proton availability.

The $R_{int}$ in mediated MFCs, as the ones used here (dual-chamber, plain carbon electrodes, Nafion membrane) are usually in the $k\Omega$ range (Table 1, *E. coli,* Ieropoulos et al., 2005). However, using high IS in combination with the halophile *H. volcanii*, we were able to obtain values compared to the ones obtained in non-mediated MFCs. But any comparison is in some way obscured by the high influence MFC design has over the majority of described performance factors, including internal resistance. To overcome this problem, we compare our archaea MFC with respect to a more widely studied *E. coli* MFC.

### 3.3. Anode and cathode potentials and pH

When the two halophilic archaea MFCs investigated here are compared, better performance could be expected to match with *N. magadii* anodes, given the highest IS of the anolyte and catholyte solutions; instead, its performance was significantly poor. This effect was also observed recently (Veer Raghavulu et al., 2009), where the effect of pH 6, 7 and 8 was assayed, finding lower $P_{max}$ at alkaline conditions (pH = 8), and higher at acidic conditions. This phenomenon was attributed to the effective extracellular $e^-$ transfer at acidophilic pH compared to basic operations, or well related to a higher activity of intracellular $e^-$ carriers. Also, Akiba et al. (1987) found 10 times less current when using alkaline microorganisms, with agree with our data (Akiba et al., 1987). But He et al. (2008) found better performance at pHs 8-10, atributed to the cathodic reaction (air cathode), that was improved by increasing pH; in the mentioned work, the electrochemical impedance spectroscopy data demonstrated that the polarization resistance of the cathode was the dominant factor limiting power output. In our set up, by using a ferricyanide cathode we assure that the cathode and cathode related reactions are not limiting the current and power production, allowing us to center our study at the anode reactions. Given the described conditions, probably the low performance of *N. magadii* MFC (when compared with *H. volcanii*) could be related to the low $H^+$ availability at pH 10 or to the biology of this archaea (metabolism, enzymes, membrane characteristics), or both.

The potentials of anode and cathode were measured with respect to a saturated Ag/AgCl reference electrode for *H. volcanii* MFC; when measured at OC or connecting RLs of 1195, 100.1 and 11.7 Ω, $E_a$ (anodic peak potential) was -198, -140, 52 and 191 m$V$; and $E_c$ (chatodic peak potential) was 300, 301, 274 and 239 mV. The data showed that anodic limitations are more important when *j* was increased, an effect that was observed previously (Jadhav and Ghangrekar, 2009). The pH values remain almost constant in the *E. coli* and *N. magadii* MFCs, but in the *H. volcanii* MFC the pH diminished during the first 24 h, reaching a value of 5.9, both at the anode and cathode compartments. During the anaerobic incubation, the accumulation of organic acids produced by the microbial strains assayed is expected, but a pH change was only evident in *H. volcanii* MFC. Probably, the buffer capacity at this MFC (lower than in *E. coli* MFC) was not enough to avoid pH changes.

### 3.4. Cyclic voltammetry studies

"Electrochemical studies are usually carried out using thoroughly cleaned glass material, mili-Q water and high purity metallic electrodes. When low-purity carbon electrodes and complex media, including living microbial cells, are electrochemically studied, the data obtained have inevitably more uncertainty, and, in some aspect, must thus be considered speculative. We evaluated our data following classical CV interpretations; comparing the results obtained using plain Toray paper electrodes at 1 mV s$^{-1}$ on plain phosphate buffer, *H. volcanii* culture media and *H. volcanii* culture, without added NR (Figure 3), it can be seen that both have a redox couple with an anodic peak ($E_a$) at ca. 128 mV, and a mid-point potential of ca.100 mV ($E_a + E_c / 2$). This couple is apparently reversible, given that the anodic and cathodic currents are ca. 2.55 µA, and the potential difference between these two peaks are ca. 58 m$V$, ($\Delta E_p = E_a - E_c$); both are reversibility criteria for one electron reaction. In this region, the plain buffer CV also shows an anodic peak at ca. 148 mV. These peaks present in all the CVs in Figure 3 may be related to material adsorbed or included at Toray paper electrodes, given that this material is relatively "dirty" when compared, e.g., with glassy carbon, a more pure electrochemical-grade electrode material. Plain buffer CV does not show any cathodic peak; also, this voltammetry does not show any other feature at negative potentials. *H. volcanii* culture also shows a second cuasi-reversible redox couple, with an $E_a$ at ca. -264 mV, mid-point potential of ca. -300 mV, and a $\Delta E_p$ of ca.71 mV. At media and buffer CV, this redox couple does not seem to be present. This quasi-reversible couple could correspond to a redox active pigments, redox active proteins or other substances produced by *H. volcanii* during growth and excreted/secreted actively by the cells, or else released from dead cells. The available information is insufficient to determinate their nature.

Besides, the sterile medium show some, less evident, redox couple at the same potentials; this means that, although in the experiments no NR was added, the current produced, at least in part, is probably assisted by chemical mediators. Other authors have found that quasi-reversible couples at CV increased with incubation time, related probably to secreted or biofilm-accumulation of electrochemical mediators (Cercado-Quezada et al., 2010). These mediators, if produced at limited quantities at *H. volcanii* MFCs, probably limited the power and current produced, as evident when NR was incorporated.

The current observed at non-mediated *H. volcanii* MFCs are probably related to this electroactive substance, or with other compound in the media; sulfate is important in the culture media (21 g L$^{-1}$), and it is well known sulfur compounds are naturally present mediators at for example sedimentary MFCs. Also Mn and Fe compounds have been proposed as mediators (Lowy et al., 2006). We do not have enough information to assign these peaks to a single redox substance, but it is evident that the current production in *H. volcanii* is assisted by external mediators. During *H. volcanii* growth, a redox soluble mediator could be produced; this strain produces carotenoid pink pigments, which are proposed as a shield against ultraviolet light. *Haloferax volcanii* contained 0.04% carotenoids of the dry wt. (Roslashnnekleiv and Liaaen-Jensen, 1995). Some of these pigments could have redox properties and could be responsible for the peaks observed with a midpoint potential of -296 mV. Endogenous mediators can be produced by biofilm growing bacteria, as demonstrated recently by Marsili (Marsili et al., 2008) where flavins secreted to the growth media for *Shewanella* are responsible at least for a part of the charge transfer to the electrode, phenomena also reported by other authors for the same bacteria (von Canstein et al., 2008).

The behavior of NR over Toray paper electrodes and high IS was also investigated; CVs performed after 10 min of $N_2$ purge are shown at Figure 4. The NR present at the *H. volcanii* growing media show a quasi-reversible behavior, with a midpoint potential of -442 mV; previous studies (Park and Zeikus, 2000) have shown a value of -524 mV (both vs. Ag/AgClsat) using fine woven graphite felt as electrodes and "normal" (ca. 100 mM) IS conditions.

**4. Conclusions**

In microbial fuel cells (MFCs), the anode contains microorganisms capable of oxidizing organic material and releasing electrons and protons. While hydrogen fuel cells use high conductive electrolytes, almost all the work published in MFCs field has been limited so far to relatively low ionic strength anodic microbial suspensions, given the concentration limits which were imposed by the physiology of the microorganisms used.

Internal resistance ($R_{int}$) is a key performance driver of fuel cells (Barbir, 2005). In mediated MFCs, ohmic resistance ($R_\Omega$) is usually the most important contributing factor to $R_{int}$. The three sources of ohmic voltage loss are: (a) resistance to ion migration within the electrolyte, (b) resistance to electron transport within the fuel cell components (electrodes, gas diffusion layer, current collectors), and (c) contact resistances (Logan, 2008). The salt concentration we used was comparable with alkaline fuel cells, were 6.6 M KOH is habitually used (Burchardt et al., 2002), allowing very low $R_{int}$ (less than 1 $\Omega$ cm$^{-2}$). The increase of NaCl is generally used at the electrolyte to improve the mass transfer of charged particles (Gil et al., 2003). The increase in the fuel cell performance a *H. volcanii* seems to be related to the increased mass transfer of charge transporters and to the increased proton availability in the cathode (pH decrease from 7.0 to 5.9 in *H. volcanni* MFC); using marine water and sediment, MFCs were reported to have better performance when compared to wastewater based ones (low IS), as reported previously (Tender et al., 2002; Bond et al., 2002). The effect of increased ionic strength was also assayed by Liu (Liu et al., 2005), where the ionic strength was increased from 100 to 400 mM, showing a noticeable power increase at high IS.

Electricity production at MFCs has been only previously linked to the metabolic activity of only very few salt-tolerant bacteria (Miller and Oremland, 2008), by using arsenate respiring bacteria isolated from moderately hypersaline Mono Lake (ML, *Bacillus selenitireducens*), and salt-saturated Searles Lake, CA (SL, strain SLAS-1); when pure culture bacteria were used, very low current was obtained for both strains, 49 and 59 µW $m^{-2}$, respectively. When the bacteria were assayed at MFC together with lake sediment, which could have some natural occurring redox mediator, significantly more power was produced. Also, in other experiments, they show that MFCs with ML sediment more power is produced (18.5 mW $m^{-2}$, salinity 90 g $L^{-1}$) than with SL sediment (1.2 µW $m^{-2}$, salinity 346 g $L^{-1}$). Although these results appear not to be consistent with our hypothesis (high power production at high salinity/IS), the highest power production at Mono lake is consistent with the following facts: microbial activities were greater in Mono Lake, *B. selenitireducens* grows faster than strain SLAS-1, and the presence of inorganic electron donors, especially sulfide, in Mono Lake sediment. Also, the presence in ML of a wider range of anaerobic bacteria capable of efficiently transferring electrons to the anode could be possible.

The present study demonstrates that more than 0.1 or 0.5 W $m^{-2}$ (non-mediated or mediated systems, respectively) ~~very elevated power densities~~ can be achieved at the high ionic strength conditions used, by reducing dramatically the resistance to ion migration within the electrolyte, permitted by the amazing physiology of the *H. volcanii* archaea. Also, we show that neutral or acidic conditions are more favorable than alkaline ones, at least using our set-up.

The new and amazing possibilities of *H. volcanii* and other extreme microbial physiology could be a key to increase the maximum current density and power obtained with MFCs. The use of added mediator allowed us to compare both ionic strength conditions; for potential applications many naturally occurring or microbially synthesized compounds can serve as electron carrier.

Co-culture of *H. volcanii* and a mediator-producing halotolerant bacterium could make the incorporation of any redox mediator unnecessary, an obvious problem when real-world applications are prospected; also it is known that waste water and sediment contain many naturally occurring substances (humic acids, iron, sulphide ions) which are known to facilitate the electricity generation (Reimers et al., 2001).

The approach used here for the first time could be a key to non-biofilm based MFC, allowing practical scaling-up. The requirement of the physical contact of the involved cells to the electrode (restricted to few microbial layers) limits the achievable density of active cells and thus the achievable power density (Reimers et al., 2001). Moreover, high volumes of brine (around 70 g $L^{-1}$ of salinity) are produced by reverse osmosis installations around the world (mainly for drinking water production). These brines are simply pumped again into the sea, but they could be used in high IS MFC waste-water depurating installations.

The high salt concentration assayed here, comparable with that used in Pt-catalyzed alkaline hydrogen fuel cells, and the use of extremophiles to cope with these conditions, are new options to increase power production and MFC scaling-up, necessary for practical applications.

**Acknowledgements**

Financial support of the National Council for Scientific and Technical Research (CONICET) and the Agency for Scientific and Technical Promotion (AGENCIA) are acknowledged.

**Figure Captions**

Figure 1. MFC Power density (open symbols) and voltage (filled symbols) as a function of current density (normalized to total geometrical electrode area). A) *H. volcanii* without neutral red (triangles) and with neutral red (circles and squares). Two independent experiments are plotted. B) *E. coli* without neutral red (triangles) and with neutral red (circles and squares). Two independent experiments are plotted.

Figure 2. MFC Power density (open symbols) and voltage (filled symbols) as a function of current density (normalized to total geometrical electrode area). *N. magadii* without neutral red at increasing absorbance, (upper triangle, down triangle, squares) and with neutral red (circles).

Figure 3. MFC Cyclic voltammetry of Toray paper anode. Plain phosphate buffer, 100 mM (solid line); *H. volcanii* sterile culture media (dash line); *H. volcanii* culture (dot line). No NR was added at any experiment, scan rate 1 mV s$^{-1}$.

Figure 4. MFC Cyclic voltammetry of Toray paper anode. *H. volcanii* culture with added NR at increasing scan rate, 50 (1), 100 (2), 200 (3) and 400 (4) mV s$^{-1}$.

**Table 1**

MFC maximum power density ($P_{max}$), and $R_{int}$ for different MFC systems. NM-MFCs are non-mediated MFCs, where direct electron transfer from bacteria to anode is postulated to occur. IS is the estimated ionic strength of the anolyte solution. ~~$R_{int}$ is strongly influenced by MFC set up, geometry, electrodes, etc, the data presented involving other authors are only partially comparable.~~

| Biocatalizer | Mediator | $R_{int}$ (Ω) | $P_{max}$ (mW m$^{-2}$) | IS (mM) | pH | References |
|---|---|---|---|---|---|---|
| *H. volcanii* | NM-MFCs | 447 ± 46 | 118.7 ± 5.4 | 2680 | 5.9 | This work |
| *H. volcanii* | NR | 66 ± 10 | 509.8 ± 36.6 | 2680 | 5.9 | This work |
| *N. magadii* | NM-MFCs | 962 | 45.7 | 3635 | 10 | This work |
| *N. magadii* | NR | 1038 | 53.8 | 3635 | 10 | This work |
| *E. coli* | NM-MFCs | 2433 ± 17 | 1.21 ± 0.08 | 92 | 7.0 | This work |
| *E. coli* | NR | 708 ± 14 | 4.71 ± 0.33 | 92 | 7.0 | This work |
| *E. coli*[a] | NR | 11,160 | 12.7 | 300 | 7.0 | Ieropoulos et al., 2005 |
| Domestic wastewater inocula | NM-MFCs | 161 | 720 | 100 | 7.0 | Liu et al., 2005 |
|  |  | 91 | 1100 | 200 |  |  |
|  |  | 83 | 1200 | 300 |  |  |
|  |  | 79 | 1330 | 400 |  |  |
| Sludge inocula | NM-MFCs | 1087 | 44.4 | 100[b] | 7.0 | Oh and Logan, 2006 |
|  |  | 625 | 75.6 | 400[b] |  |  |
| *Geobacter metallireducens*[c] | NM-MFCs | 19,920 | 2.2 | 158 | 7.0 | Min et al., 2005 |
|  |  | 1286 | 40 | 158 |  |  |

[a]Recalculated data considering anode geometrical area (10 cm$^2$); the indicated $P_{max}$ correspond to a RL of 10 kΩ.

[b]IS was estimated from data presented by the authors.

[c]Data obtained from 2 types of MFC, salt bridge ($R_{int}$ = 19,920 Ω) and membrane based MFC with lower $R_{int}$.

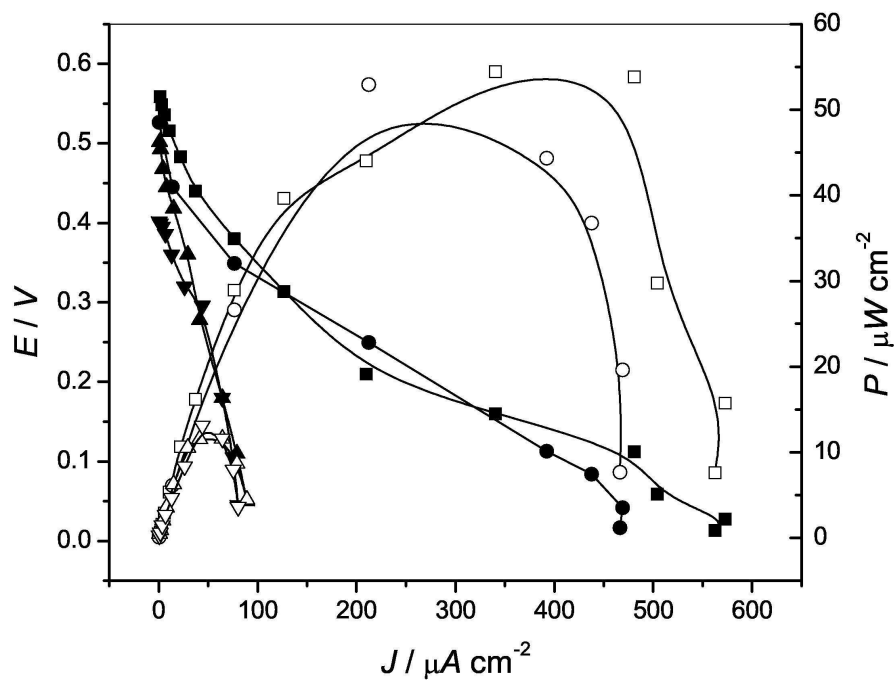

Figure 1. MFC Power density (open symbols) and voltage (filled symbols) as a function of current density (normalized to total geometrical electrode area). A) H. volcanii without neutral red (triangles) and with neutral red (circles and squares). Two independent experiments are plotted. B) E. coli without neutral red (triangles) and with neutral red (circles and squares). Two independent experiments are plotted.
104x81mm (600 x 600 DPI)

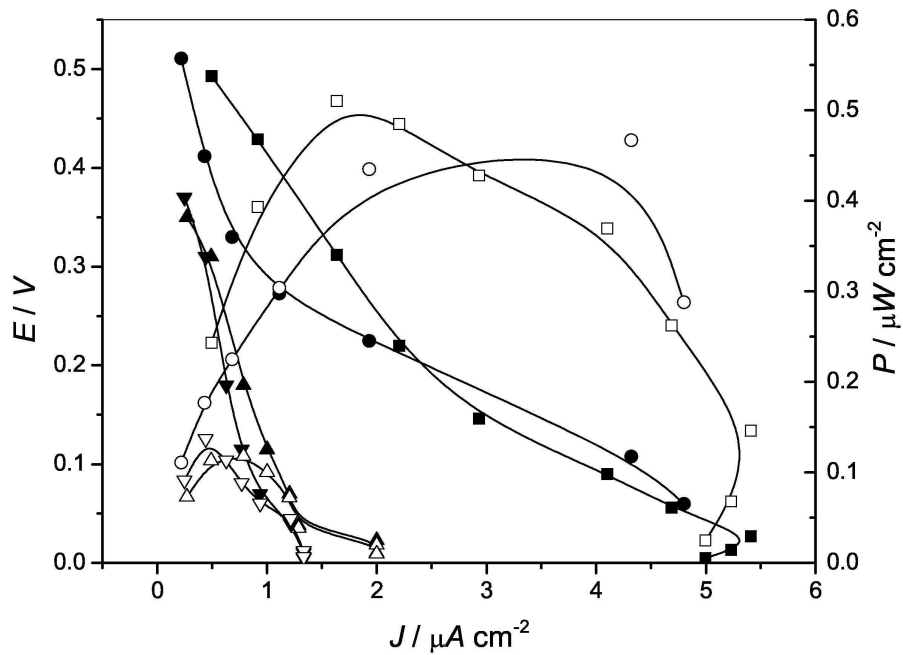

Figure 1. MFC Power density (open symbols) and voltage (filled symbols) as a function of current density (normalized to total geometrical electrode area). A) H. volcanii without neutral red (triangles) and with neutral red (circles and squares). Two independent experiments are plotted. B) E. coli without neutral red (triangles) and with neutral red (circles and squares). Two independent experiments are plotted.
111x83mm (600 x 600 DPI)

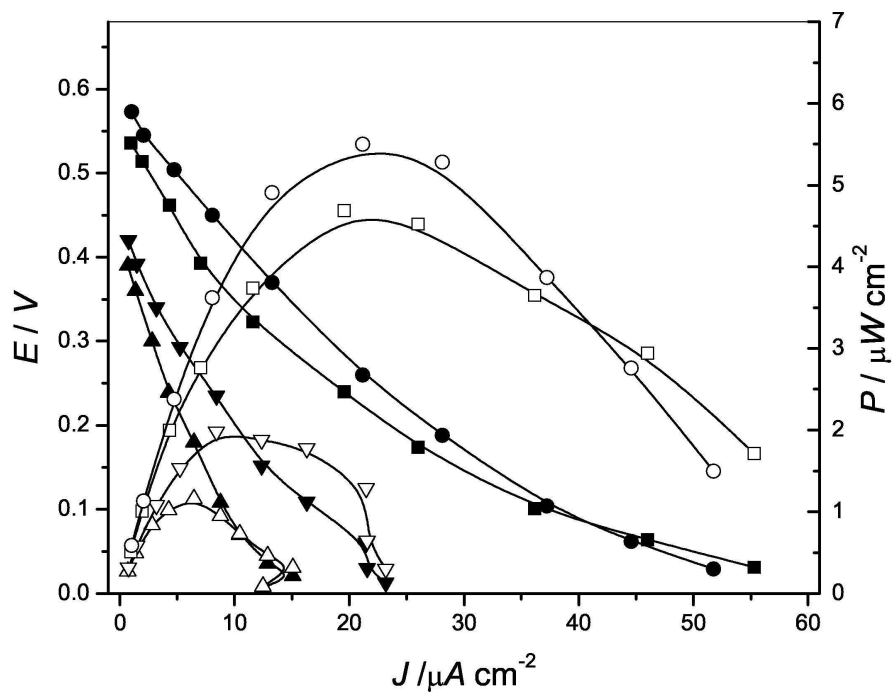

Figure 2. MFC Power density (open symbols) and voltage (filled symbols) as a function of current density (normalized to total geometrical electrode area). N. magadii without neutral red at increasing absorbance, (upper triangle, down triangle, squares) and with neutral red (circles).
101x81mm (600 x 600 DPI)

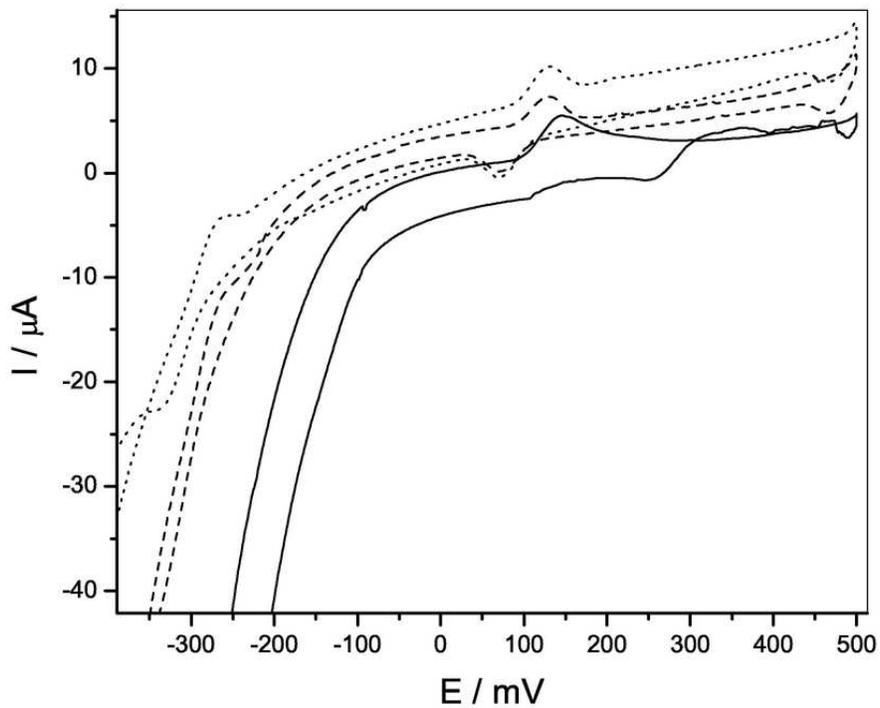

78x64mm (300 x 300 DPI)

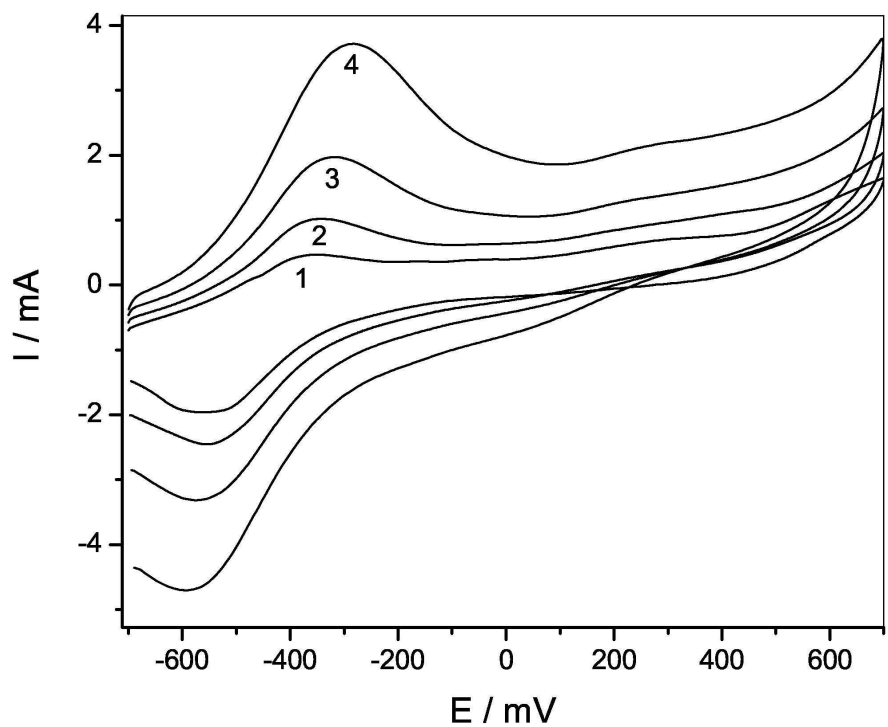

Figure 4. MFC Cyclic voltammetry of Toray paper anode. H. volcanii culture with added NR at increasing scan rate, 50 (1), 100 (2), 200 (3) and 400 (4) mV s-1.
94x80mm (600 x 600 DPI)